# A Dynamical Analysis of the Dust Tail of Comet C/1995 O1 (Hale-Bopp) at High Heliocentric Distances

Emily A. Kramer[a],*, Yanga R. Fernandez[a], Carey M. Lisse[b], Michael S. Kelley[c], Laura M. Woodney[d]

[a]Dept. of Physics, Univ. of Central Florida, 4000 Central Florida Blvd., Orlando, FL 32816-2385, USA
[b]Applied Physics Laboratory, Johns Hopkins Univ., 11100 Johns Hopkins Rd., Laurel, MD 20723, USA
[c]Dept. of Astronomy, Univ. of Maryland, College Park, MD 20742-2421, USA
[d]Dept. of Physics, California State Univ., 5500 University Parkway, San Bernardino, CA 92407-2318, USA

* Corresponding Author. Email address: ekramer@alum.mit.edu (E.A. Kramer)

## **Abstract**

Comet C/1995 O1 (Hale-Bopp) has provided an unprecedented opportunity to observe a bright comet over a wide range of heliocentric distances. We present here Spitzer Space Telescope observations of Hale-Bopp from 2005 and 2008 that show a distinct coma and tail, the presence of which is uncommon given its large heliocentric distance (21.6 AU and 27.2 AU, respectively). The morphology of the dust is compared to dynamical models to understand the activity of the comet. Our analysis shows that the shape of Hale-Bopp's dust tail in these images cannot be explained using the usual Finson-Probstein (solar gravity + solar radiation pressure) dynamical model. Several alternative explanations are explored. The analysis suggests that the most likely cause of the discrepancy is that the dust is being charged by the solar wind, then being affected by the interplanetary magnetic field via the Lorentz force. Though this effect has been explored previously, if correct, this seems to be the first time that the Lorentz force has been required to model a cometary dust tail. The analysis also suggests that Hale-Bopp was actively emitting particles when these images were taken, and the tail characteristics changed between observations.

## Section 1: Introduction, Background and Motivation

A fundamental goal of astronomy is to understand the formation of Earth and our Solar System. Since it is impossible to observe our Solar System in its early stages, the next best thing is to study bodies that have been relatively preserved over the age of the Solar System. Comets have remained relatively unchanged since their birth as icy planetesimals in the early days of the Solar System, so they are a prime example of objects that can help us to understand the mechanics and chemistry of the planetary formation era. However, comets are not pristine; they have undergone physical and chemical alterations via various processes such as collisions (Farinella and Davis, 1996) and insolation (Huebner and Benkhoff, 1999). These changes require that we study cometary evolution to fully understand what the primordial comets were like.

### Section 1.1: Background

Hale-Bopp was discovered in July of 1995, when it was inbound at a heliocentric distance of 7.2 AU. At this large distance it had already developed a coma, suggesting that activity was occurring on the nucleus at that time. It is rare to see such strong cometary activity beyond ~5 AU, so it became immediately clear that Hale-Bopp's behavior was different from that of other comets. The activity at this point was at least partially driven by the sublimation of CO (Biver et al. 1996), and transitioned to water driven activity at around 3 AU (Biver et al. 1997). When it reached its perihelion at 0.9 AU in 1997, its mass-loss rate was very high, $(1.8 \pm 0.4) \times 10^6$ kg s$^{-1}$, and the total mass loss while it was inside 3 AU was estimated to be ~ $3 \times 10^{13}$ kg (Lisse et al., 1997; Jewitt and Matthews, 1999). These extreme values were likely due to the fact that the nucleus is large ($30 \pm 10$ km in radius and ~$10^{15}$ kg in mass) (Fernández et al., 2002) and its dust to gas ratio was 5:1 (compared to the canonical ratio of 1:1), meaning that it ejected a relatively large amount of dust (Jewitt and Matthews, 1999). Assuming an average nuclear radius of 30 km and a density of 1 g cm$^{-3}$, this results in an average loss of surface material to a depth on the order of meters.

### Section 1.2: Motivation

While more and more comets are recognized to be active at ~5 AU (e.g. Fernandez et al. 2013, Epifani et al. 2009), activity at even greater distances is rarely seen. Whether this is due to detector sensitivity limits or a true lack of activity is an open question. Observations collected when a comet is far from the Sun are more likely to yield data that can be used to characterize the nucleus. One such study was done by gathering observations of 21 comets using the Hubble Space Telescope and the Keck II telescope, with heliocentric distances out to 29.5 AU (Meech et al., 2004b). This study focused on objects that were inactive at the time of observation, as is expected for most comets at large heliocentric distances. Distant observations have been made on other comets, including Halley, which was observed to be active at 14.3 AU, but inactive at 28.1 AU (Hainaut et al. 2004).

Thus, while observations of inactive comets had been made at large heliocentric distances, the prospect of obtaining the first mid-infrared imaging and photometry of a truly distant active comet motivated us to observe Hale-Bopp with the Spitzer Space Telescope in 2005 and 2008. The 2008 observations show the most distant cometary activity ever recorded, when Hale-Bopp was at 27.2 AU. This is significant, since it allows cometary activity to be studied in an entirely new heliocentric regime. Other researchers have also observed distant activity on Hale-Bopp at 25.7 AU (Szabo et al. 2008), but more recent observations have suggested that activity has likely ceased by 30.7 AU (Szabo et al. 2011). These observations were obtained in the visible wavelength regime, and are therefore complementary to the infrared data presented here.

The remainder of the paper has the following structure: Section 2 describes the images and data reduction, and Section 3 describes the analysis of the nucleus and general tail structure. Section 4 goes in-depth into the different dynamical models used, and Section 5 gives the main conclusions of the paper and a discussion of the results.

## Section 2: The Data

The data for this project were collected using the Spitzer Space Telescope (Werner et al. 2004). The instrument used was the Multiband Imaging Photometer for Spitzer (MIPS) (Rieke et al. 2004), operating in 24μm photometry mode. This mode collected data in sets of 12 dithered images to reduce effects from bad pixels. The MIPS Si:As detector has 128x128 pixels with a ~5'x5' field of view, with a pixel size of 2.59 arcseconds/pixel.

The data were collected in two campaigns, in May 2005 and August/September 2008. Details about the observations are listed in Table 1. Note that even though the data were collected over two years apart, the comet's large heliocentric distance (>20 AU) meant that there was only a small change in Hale-Bopp's phase angle with respect to Spitzer. This allowed for better comparison of the data between the two epochs than if the difference in phase angle had been large. The temperatures listed in Table 1 come from modeling the thermal behavior of cometary dust as a spherical isothermal black body, assuming the equation $T = 278K/\sqrt{r}$, where $r$ is in units of AU (Reach 1988). This is a good first-order approximation for the behavior of Hale-Bopp, since cometary dust has low albedo and generally high emissivity. Thus the peak of Hale-Bopp's spectral energy distribution was near 48 μm in 2005 and 53 μm in 2008. This implies that the MIPS detector was collecting thermal emission from the Wien's Law side of Hale-Bopp spectral energy distribution.

The dithered images from each night were combined into a single image using the Mosaicker and Point source Extractor (MOPEX), a program created by the Spitzer Science Team to reduce and analyze data from the Spitzer Space Telescope. This software allows the user to stack dithered images gathered by Spitzer instruments to create a single mosaicked image for each set of observations (Makovoz and Khan 2005). However, the stacking is done relative to the stars rather than in the frame of reference of the target object. If an object is near to the Sun, and therefore moving quickly relative to the stars, this can have the effect of smearing the object across the image. Fortunately, for the case of Hale-Bopp, this effect is nearly negligible. In 2005 the total projected movement during the entire integration time was 2.7 arcseconds, or ~1 MIPS pixel. For 2008, the total projected motion is 6.9 arcseconds (due to the longer total integration time), or ~2.3 MIPS pixels. The structure of the tail that is being studied here is >> 2 pixels in all dimensions, therefore this motion does not affect the analysis of the dust tail. However, it will significantly degrade the quality of any nucleus analysis. It must be noted that the final mosaicked image has a pixel size of 2.45 arcseconds per pixel, due to the image stacking algorithms.

The final images are shown in Figure 1.  In the 2005 images, an extended dust tail is clearly seen. In 2008, Hale-Bopp was in a relatively crowded star field, making it difficult to determine the extent of the tail. To ameliorate the confusion from the stars, the data from 1 September 2008 were subtracted from the data for 31 August 2008, using it as a shadow image that removes the background stars. The difference leaves a negative image, which can be seen in Figure 1b.  Artifacts from stars that did not perfectly match up are still visible, but none are near enough to the tail to affect the results of the analysis.

## Section 3: Analysis

### Section 3.1: Analysis of the Nucleus
Our first goal is to determine if the flux in the head of the comet was emitted by a bare nucleus (giving it a radial profile akin to a reference star's profile) or a cloud of dust surrounding the nucleus (producing a wider radial profile). A bare nucleus would indicate that activity has ceased on the comet, while the presence of a coma suggests that activity is still taking place.  We compare the radial profile of the comet's nucleus to the radial profile of a star in the frame.  This comparison was only done for 2005, since the 2008 shadow images had the field stars removed, and the non-shadow images had stars too close to the nucleus for the exercise to be meaningful. The results of this comparison are shown in Figure 2.  Hale-Bopp's extended emission is shown by its wider radial profile than that of the field star.

Assuming that the nucleus is a bare (non-active) blackbody point source, the expected flux from the nucleus is 0.2-0.8 mJy in 2005 and 0.05-0.2 mJy in 2008.  The range of values is due to the fact that Hale-Bopp's effective nuclear radius is uncertain by a factor of 2 (Toth and Lisse, 2006).  Photometric analysis was done on the 2005 data and the positive image of the 2008 data for a range of aperture sizes.  For 2005, the aperture integrated flux ranged from 22 mJy for a 2-pixel radius aperture to 468 mJy for a 5-pixel aperture. For 2008, the central pixel flux ranged from 3.2–60 mJy for the same range of apertures.  In both epochs, the observed flux is approximately three orders of magnitude brighter than would be expected for a bare nucleus (Fernández et al. 2000), again indicating that there is a significant coma present.

### Section 3.2: Analysis of the General Tail Structure
Next, the morphology of Hale-Bopp's tail was characterized in each mosaicked image. The images were first rotated such that the tail was oriented approximately along the horizontal axis.  A Gaussian was then fit along each vertical pixel column through the tail.  Figure 3 shows the results of this analysis.  Note the small orbit plane angle (< 3°) and low phase angle (also < 3°) for both epochs introduce severe projection effects. The tail is likely greatly foreshortened, so a good deal of structural features may be lost.

The first point to note is that there is no debris trail present along Hale-Bopp's orbit plane. A 2007 study of 34 Jupiter-family comets observed by Spitzer showed debris trails to be present in at least 27 of them (Reach et al. 2007). These observed trails are comprised of 100 μm to mm-sized dust grains, and can be thought of as small chunks of the comet that break off during its orbit, and then orbit on nearly the same path as the original body. This phenomenon is not unique to Jupiter-family comets. Cometary dust trails from Halley-family comets such as 1P/Halley and 109P/Swift-Tuttle are known to be the cause of meteor showers on Earth. The absence of a detectable trail in the Hale-Bopp images suggests that few large, slow grains were released earlier during the comet's orbital period.

Note that the shape of the tail is dramatically different in 2005 and 2008. In 2005, the tail seems to have more curvature, and is wider even when the difference in distance from Spitzer is taken into account. The tail seen in the 2008 data is much narrower, and straighter. The projected width of the tail in 2005 is approximately $9.1 \times 10^5$ km, while the projected width in 2008 is $5.1 \times 10^5$ km. The curved structure in the tail seems to be real, and not due to background stars compromising the fitting, since it is present along the entire length of the tail. The dust emission was well fit out to a projected distance of approximately $2.3 \times 10^6$ km in 2005 and $2.4 \times 10^6$ km in 2008. The tail could not be fit well beyond that distance. The differences in tail width and curvature could be due to differences in viewing geometry, since the orbit plane angle changes by a factor of 2.75 between the two observations (see Table 1).

Tail orientation, measured in terms of position angle, or orientation relative to the fixed coordinate system, is diagnostic of the origin of cometary tails. Examination of Figure 1 reveals that the position angle of the tail changed approximately 70° between 2005 and 2008. This cannot be explained by changes in viewing geometry of the system, since Table 1 shows that the phase angle relative to Spitzer ($\alpha$) hardly changed at all between observations. In fact, it appears that the tail more closely follows the anti-sunward direction in 2008 than in 2005. However, the angle between the tail and the extended Sun-comet vector changes by only about 5° between 2005 and 2008, with the tail closer to the anti-sunward direction in 2008.

Additionally, the tail fades more than would be expected if the cometary activity had remained constant. This can be quantified in the following way: if we assume that the emission is from isothermal spheres, the estimated band radiance can be calculated. Spitzer's 24μm camera has a spectral range of 21-27μm. As described above, the temperature of Hale-Bopp's dust is estimated to be 60K in 2005 and 54K in 2008. This yields a band radiance of 4.21 mW/m$^2$/sr in 2005 and 1.42 mW/m$^2$/sr in 2008, giving a ratio of 0.338. Thus, all other factors being equal, it would be expected that the dust seen within the same angular aperture in 2008 would be 33.8% as bright as the dust seen in 2005. To measure this in the data, a section of the tail was selected from the 2005 data that was free from background stars, and far enough from the nucleus as to not be affected by the coma. The projected distance to that location was calculated, and then transformed into the pixel scale for the 2008 data, thereby determining where the aperture should be located. The section of tail measured was 10x5 pixels for each image. The flux in the 2005 tail box was measured to be 59 mJy and 10 mJy in 2008. This yields a ratio of 0.177, which is about half as bright as the tail would be if we were looking at the same dust grains in both epochs. Instead, it suggests that activity is dropping off as heliocentric distance increases, as one would expect.

The differences in tail shape, orientation and brightness between 2005 and 2008 lead us to conclude that the dust we see in the two images is new, that is, it has been recently emitted while Hale-Bopp was very far from the Sun. This is significant since it is rare to see activity so far after the comet has left the water sublimation zone, and even rarer outside the more distant $CO_2$ sublimation zone. Observation of activity at such large heliocentric distances indicates that something other than water ice sublimation, most likely amorphous ice crystallization, is causing the activity (Meech and Svoren 2004). In a study of the activity of comet Halley at large heliocentric distances, Flammer et al. (1986) proposed that brightness variations were caused by small particles becoming electrically charged and then carried away by the high-speed solar wind. While the mechanism for lifting dust particles off of Hale-Bopp's surface at large heliocentric distances is an interesting question, it is beyond the scope of the current paper. The modeling techniques described below are not dependent on the source of the activity.

## Section 4: Modeling the Dust Tail

### Section 4.1: The Finson-Probstein Model

A widely used model for cometary dust tails the Finson-Probstein model (Finson and Probstein 1968). This model assumes that once cometary dust particles leave the surface, their motion is governed by two forces: solar gravity and solar radiation pressure. The particle motion can then be parameterized using the ratio of these two forces, called $\beta$:

$$\beta = F_{rad}/F_{grav} \qquad [1]$$

In physical units, this gives the ratio:

$$\beta = \frac{C Q_{pr}}{\rho_d a} \qquad [2]$$

where $Q_{pr}$ is the scattering efficiency for radiation pressure, $\rho_d$ is the mass density of the particle in g cm$^{-3}$, $a$ is the particle radius in cm, and the factor of $C$ = 5.76 x 10$^{-5}$ g cm$^{-2}$ comes from multiplying all the constant values (Finson and Probstein 1968). Thus $\beta$ depends on the inverse of particle radius; i.e. for small grains, $\beta$ is larger, meaning the radiation pressure pushing the particles outwards has a larger effect than the gravitational force pulling them inwards. $\beta$ is incorporated into the equation of motion in the following way:

$$\ddot{\vec{x}} + (1-\beta)\frac{GM_s}{|\vec{x}|^3}\vec{x} = 0 \qquad [3]$$

where $G$ is the universal gravitational constant, $M_s$ is the mass of the Sun, and $\vec{x}$ is the vector position of the object. This is a simple equation of motion that can then be integrated to track the motion of particles with different $\beta$ values.

The computations were carried out by creating a numerical integrator (based on the work of Lisse et al., 1998) in the language *Python*. The software takes in a set of $\beta$ values, integrates the motion of the dust particles over the designated time interval, and returns a set of curves (i.e. syndynes) that show the positions of the grains. Each syndyne corresponds to dust with a particular $\beta$ that was released continuously from some given time ago up to the time of the image. Since the forces on particles of different $\beta$ are different, the syndynes will tend to fan out in the comet's orbital plane. If the data are well modeled by the syndynes, the curves will span the width of the dust tail when overplotted on the data image. In the general case, the relative velocity of the grains from the comet's surface can be included into the integration, giving individual $\beta$ curves a spread of some finite width. In this case, we are concerned with general shape matching, so the particles were given no initial velocity relative to the nucleus.

The results of the Finson-Probstein analysis are shown in Figures 4 and 5. For both the 2005 and the 2008 data, a range of $\beta$ values (10, 3, 1, 0.3, and 0.1) were used to generate the syndynes shown. It is immediately obvious that this model does not fit the data well, since the curves do not span the width of the dust tail. In the 2008 data (Figure 5) it may seem that the $\beta$=10 value works well, but a $\beta$ that high is actually unphysical (Burns, Lamy and Soter, 1979).

## Section 4.2: The Rocket Force

Another phenomenon that proved fruitful with modeling the dust of another comet (73P/Schwassmann-Wachmann 3) (Reach et al. 2009) is the so-called "rocket force". This effect is caused by sublimation of surface ice on the day side of ejected grains, which causes them to move in the anti-sunward direction at greater than expected velocities. Similar to $\beta$ in the Finson-Probstien model, the rocket force can be expressed as a ratio of forces:

$$\alpha \equiv \frac{F_{rocket}}{F_{gravity}} = \frac{3\,\mu\,Z\,v_{ice}f_{ice}}{4\,G\,M_s\,\rho\,a} \tag{4}$$

where $\mu$ is the molecular weight of the ice, $Z$ is the ice sublimation rate, $v_{ice}$ is the velocity of expansion of the sublimating ice, $f_{ice}$ is the fraction of the icy surface that is exposed, $G$ is the universal gravitational constant, $M_s$ is the mass of the sun, $\rho$ is the grain mass density, and $a$ is the grain radius.

The $\alpha$ value here can be thought of as analogous to the $\beta$ parameter used in the classic Finson-Probstein model: it is a ratio of a force that pushes the grains outwards over the gravitational force of the Sun pulling the grains back inwards. It is mathematically equivalent to $\beta$ if the rocket force is radial. Thus, running the models for different $\alpha$ values would give the same results as running it for the same $\beta$ values. Additionally, by definition, the rocket force only works on grains with day-night temperature anisotropies. This idea is hard to justify for the small, cold grains that are present in Hale-Bopp's tail. With 73P, Reach et al. (2009) use millimeter or centimeter sized grains, which are likely much larger than the grain sizes seen in the Spitzer 24μm data. Therefore, we conclude that this method is not useful for our situation.

## Section 4.3: Jets

When strong jets are present on a comet, it is typical for the coma and tail shape to differ dramatically from the shape expected by the isotropic emission that is assumed by the Finson-Probstein model (e.g., Schleicher and Farnham 2004). This is due to the emission coming off of specific parts of the comet's surface, rather than slowly from all over the comet's surface. In order to determine whether this was the cause of the discrepancy between the model and the data, a 3D visualization technique was used. A mock (spherical) comet was created, with vectors to point to the Sun direction, the coordinate axes, and Hale-Bopp's pole direction (Licandro et al. 1998; Farnham et al. 1999). We then determined the locus of points on the mock comet where the jet would need to be for the observed emission to come from it for 2005 and 2008 separately. The analysis shows that the jet would need to be in a different location on Hale-Bopp's surface in 2005 and 2008. Since the temperature has changed little between the two observations, and since the sub-solar latitude on the comet has likewise changed little, it is unlikely that we are seeing seasonal effects. Thus it is unlikely that one jet would shut off and another turn on in the 3 years between the data sets. Additionally, the jets would need to be active for only a short time during Hale-Bopp's rotation period, or else a spiral structure would appear in the images. The spiral structure would not appear if the jet was located close to the rotational axis pole, but the analysis has shown that the jet(s) could not be close to either pole. The conditions required to produce the tail structure seen in this data using jet modeling (i.e. different jet locations in 2005 and 2008 and jets that are only active for part of the comet's rotation period) are highly contrived and unlikely to occur naturally. Thus, this method also does not solve the problem.

## Section 4.4: The Lorentz Force

With none of the previously accepted modeling techniques satisfactorily fitting the data, an alternative approach was next considered. The interaction between cometary gas tails and the solar wind has been the subject of numerous studies (e.g. Alfvén, 1957; Brandt, 1968), and the interaction between the solar wind and the ions within the gas tail is known to be the reason for their shape (Biermann, Brosowski & Schmidt, 1967). The use of this technique has been investigated by a number of previous studies (e.g. Finson and Probstein 1968; Burns, Lamy & Soter, 1979; Wallis and Hassan, 1983; Lisse et al. 1998), and was detailed for the case of the Rosetta mission by Mendis and Horanyi (2013). In particular, Wallis and Hassan (1983) demonstrate that for sub-micron sized dust particles, the Lorentz force is stronger than radiation pressure in the coma, and thereby substantially changes particle trajectories within the coma. However, these investigations have been primarily concerned with relatively small-scale structures for comets that were near the ecliptic plane. Below, we demonstrate that for the case of Hale-Bopp at large heliocentric distances, the charging of cometary dust grains becomes important to explain the shape of the dust tail.

The solar magnetic field acts as a force upon the particles via the Lorentz force. In the general case, the Lorentz force is defined as:

$$F_L = q(\vec{v} \times \vec{B}) \qquad [5]$$

where $q$ is the charge on the particle, $\vec{v}$ is the relative velocity between the particle and the field carrying the charges and $\vec{B}$ is the vector strength of the magnetic field. With the addition of the Lorentz force, the equation of motion for the particles then becomes:

$$\ddot{\vec{x}} + (1-\beta)\frac{G M_s}{|\vec{x}|^3}\vec{x} - \frac{q}{m}\left(\left(\dot{\vec{x}} - \vec{v}_{sw}\right) \times \vec{B}\right) = 0 \qquad [6]$$

where $q/m$ is the charge-to-mass ratio for the particles and $\vec{v}_{sw}$ is the vector velocity of the solar wind. The magnetic field $\vec{B}$ can be broken down into the radial ($B_r$), azimuthal ($B_\Phi$) and normal ($B_\theta$) components of the Parker Field as a function of heliographic latitude ($\beta_{hg}$) and heliocentric distance ($r$):

$$B_r = \pm B_{r,0}\left(\frac{r_0}{r}\right)^2 \qquad [7]$$

$$B_\Phi = \pm B_{\Phi,0}\frac{r_0}{r}\cos(\beta_{hg}) \qquad [8]$$

$$B_\theta = 0 \qquad [9]$$

where $B_{r,0} = B_{\Phi,0} = 3\text{nT}$ are the radial and azimuthal field strength, respectively, at $r_0 = 1$ AU (Landgraf 2000).

The Lorentz force acting on cometary dust particles is caused by the solar wind, which is essentially radial and moves at a speed of $v_{sw}$. The speed of the solar wind at high heliographic latitudes (>|28|°) has been measured by the Ulysses spacecraft to be ~750 km/s (Smith et al. 2000). Since Hale-Bopp was at a heliographic latitude of -83.1° in 2005 and -84.9° in 2008, we can use this measurement. Hale-Bopp's heliocentric speed in 2005 and 2008 was 8.6 km s$^{-1}$ and 7.6 km s$^{-1}$, respectively. Since this is much less than the solar wind speed, we can take $|\vec{v}|$ to be essentially $v_{sw}$. Thus, when we take the cross product in Equation 5 only the azimuthal component of the magnetic field survives, giving:

$$F_L = \pm \frac{q}{m} B_{\Phi,0} \frac{r_0}{r} \cos(\beta_{hg}) v_{sw} m \qquad [10]$$

where $m$ is the particle mass. The parameter $q/m$ can be calculated by combining the formula for the calculation of a charge on a particle

$$q = 4\pi\varepsilon_0 V a \qquad [12]$$

where $V$ is the surface potential on the particle, $\varepsilon_0$ is the permittivity of free space, and $a$ is the particle radius, with the formula for the mass of a particle

$$m = \frac{4}{3}\pi a^3 \rho_d. \qquad [13]$$

Using the formula for β from Equation 2, we can then obtain a formula for $q/m$

$$\frac{q}{m} = \frac{12\varepsilon_0 V}{C^2} \beta^2 \frac{\rho_d}{Q_{pr}^2} \qquad [14]$$

recalling that $C$ = 5.76 x 10$^{-5}$ g cm$^{-2}$. Since $q/m$ goes as $\beta^2$, and thus a$^{-2}$, the Lorentz force is largest for smaller particles. The dust grains acquire a surface potential of $V$ = +5V by the photoelectric effect from solar ultraviolet radiation. Note that this formula has two free parameters ($\rho_d$, and $Q_{pr}$) for each β value, so this modeling technique can be used to investigate the density and/or scattering efficiency of the dust tail particles. Table 2 shows the $q/m$ values that are obtained for a sample particle with β = 0.3 and a range of density and scattering efficiency values. For a spherical β = 0.3 particle, this gives a few thousand charges on the grain.

We added the Lorentz force to the code that was originally used for the Finson-Probstein model, and reran the simulations with the same β values as used before and canonical density ($\rho_d$=1000 kg/m$^3$) and scattering efficiency ($Q_{pr}$=1.0) (Burns, Lamy, and Soter, 1979) values to best show the effect of the Lorentz force. The results are shown in Figures 6 and 7. It is clear that this method does a much better job of modeling the data, as the syndynes now fully span the width of Hale-Bopp's dust tail for both 2005 and 2008. By closely examining the figures, one can see that the β = 0.3 syndyne most closely matches the tail in both observations, with the 3-year syndyne most closely matching in 2005 and the 5-year syndyne most closely matching in 2008. Depending on the choice of density and scatting efficiency used to calculate the particle size, this gives a particle radius of 0.48 to 5.76 μm. We can therefore limit the need for the non-physical parameters that were needed to get even mediocre results using the standard Finson-Probstein method.

We have also explored the use of a range of density and scattering efficiency parameters in an effort to explore the effect of these parameters on the models. We ran the models for the parameter space spanning $Q_{pr}$= [0.5, 1.0, 1.5] and $\rho_d$= [500, 1000, 2000] kg/m$^3$. Figures 8 and 9 show the resulting figures, which suggest that density and scattering efficiency tend to rotate the syndynes in opposite directions. None of the alternative cases was a significantly better fit to the data in both 2005 and 2008, and thus there is no justification to select a density or scattering efficiency value that is different that what would be expected from the literature.

It is important to note that the values we used in the model for the strength of the magnetic field ($B_P$) and the speed of the solar wind ($v_{sw}$) are entirely typical. That is, in order for this method to model the data well, we did not need to set $B_{\Phi,0}$ or $v_{sw}$ to unusual values. The expected, average numbers worked well in our models. This suggests that invoking the Lorentz force is a reasonable hypothesis to explain the deviation from the standard model.

It should be mentioned that any interaction between the stellar magnetic field and the nucleus of the comet would not affect these results. This is due to the fact that the size scale for this interaction ($\sim 10^3$ km) is significantly smaller than the size scale for the tail seen in these images ($\sim 10^6$ km, projected) (Ip 2004).

### Section 5: Conclusions and Discussion

There are two main results that have arisen from the analysis of these data:

(1) **Hale-Bopp was active at these large heliocentric distances**, and the time scale for evolution of the tail is short enough that significant changes could be observed over the course of 3 years.

(2) **We infer from our analysis of the data that the Lorentz force likely plays a significant role in the motion of Hale-Bopp's dust at this point in its orbit,** and other standard models cannot fully explain the dust dynamics. **This is the first time that the Lorentz force has been needed to explain the observed morphology of a cometary dust tail.** Based on the models which incorporate the Lorentz force, the tail is comprised of $\beta = 0.3$ particles, which corresponds to a particle radius of $\sim 0.5$ - 5 $\mu$m.

The natural question that arises from this analysis is: "if the Lorentz force has such a dramatic effect on Hale-Bopp's dust tail as seen in these images, why has this not been observed in other comets, or even in other observations of Hale-Bopp?" The reason for this is best explained by examining the equations that govern the motion of the particles. As shown in Section 4.1, the equation for the radiation pressure force and gravity depends on the distance from the Sun as $1/r^2$, while the equation for the Lorentz force, shown in 4.4, depends on heliocentric distance as $1/r$ times the heliographic latitude as $\cos(\beta_{hg})$. Thus, for comets at large heliocentric distances and low heliographic latitudes, the Lorentz force should play a significant role in the dynamics of the dust. In Hale-Bopp's case, the heliographic latitude increases rapidly due to its high inclination, thereby causing the effect of the Lorentz force to also drop off quickly. However, it does not drop off as quickly as either radiation pressure force or gravity, thus at high heliocentric distances it becomes important. The strength of each of the three forces as a function of Hale-Bopp's orbit is shown in Figure 10. The images used in this analysis were gathered when Hale-Bopp was at $r = 21.6$ and 27.2 AU, which is significantly more distant than the vast majority of cometary observations. In fact, at the time these observations were gathered, they were the largest heliocentric distance at which cometary activity had been ever observed.

Note that while the presence of a coma or dust tail does not necessarily indicate that activity is occurring when the image was captured, the fact that the models indicate that the dust grains are of size order 1μm, and the fact that we see intrinsic morphology and photometric changes to the tail suggest that activity was occurring between the time when the two data sets were observed. This is due to the fact that small dust grains would have moved quickly away from the comet nucleus. If 1μm sized dust grains had been released when the comet was at, say, perihelion, they would be so far away from the nucleus as to be far outside the field of view of the images. Additionally, if the grains were large, the images from 2005 and 2008 would look more similar, since the grains would be moving relatively slowly and wouldn't change their position relative to the nucleus very much over the 3.3-year gap.

In the future, it would be useful to continue to observe Hale-Bopp in infrared and optical wavelengths until it is no longer observable. We might then be able to detect when it "turns off" (the activity stops), which would give insight into which volatiles are most important for generating dust activity. Such work has been done by Szabó (2012) with the Herschel Space Telescope and the Very Large Telescope, which suggests that the nucleus had ceased to produce new gas and dust before their observations in mid-2010 (Szabó et al. 2012). Continued deep imaging of the tail would also provide further tests for the Lorentz force hypothesis.

It would also be beneficial to observe additional comets that show activity at large heliocentric distances, since their dust tails may show the effect of the Lorentz force. If this could be done for a number of comets at a range of heliographic latitudes and heliocentric distances, this could be a new way to probe the structure of the solar magnetic field.

### **Acknowledgments**

We thank Mihaly Horanyi and an anonymous reviewer for their comments that helped to improve this paper. This work is based on observations made with the Spitzer Space Telescope, which is operated by the Jet Propulsion Laboratory, California Institute of Technology under a contract with NASA. Support for this work was provided by NASA through an award issued by JPL/Caltech, RSA #1348873. Support for this work was also provided by the National Science Foundation through grant AST-0808004, and by NASA through grant NNX-09AB446 through the Planetary Astronomy Program. EK was also funded by NASA Earth and Space Sciences Fellowship NNX-12AN88H.

**Tables**

**Table 1:** Observation Details

| Observation Date | r (AU) | Δ (AU) | α (°) | Orbit Plane Angle (°) | Number of Images | Total Exposure Time (hours) | Blackbody Equilibrium Temperature (K) |
|---|---|---|---|---|---|---|---|
| 19 May 2005 | 21.6 | 21.4 | 2.7 | -2.2 | 144 | 0.42 | 60 |
| 31 Aug 2008 | 27.2 | 26.9 | 2.1 | 0.8 | 432 | 1.26 | 54 |
| 1 Sep 2008 | 27.2 | 26.9 | 2.1 | 0.9 | 432 | 1.26 | 54 |

r: Heliocentric distance; Δ: Spitzer distance, α: Spitzer phase; T calculated from $T = 278/\sqrt{r}$

**Table 2:** Calculated q/m values for $\beta = 0.3$ in C/kg

| | | $Q_{pr}$ Values | | |
|---|---|---|---|---|
| | | **0.5** | **1.0** | **1.5** |
| **Density (kg/m³)** | **500.** | $2.88 \times 10^{-1}$ | $7.20 \times 10^{-2}$ | $3.20 \times 10^{-2}$ |
| | **1000.** | $5.76 \times 10^{-1}$ | $1.44 \times 10^{-1}$ | $6.40 \times 10^{-2}$ |
| | **1500.** | $8.65 \times 10^{-1}$ | $2.16 \times 10^{-1}$ | $9.61 \times 10^{-2}$ |
| | **2000.** | 1.15 | $2.88 \times 10^{-1}$ | $1.28 \times 10^{-1}$ |

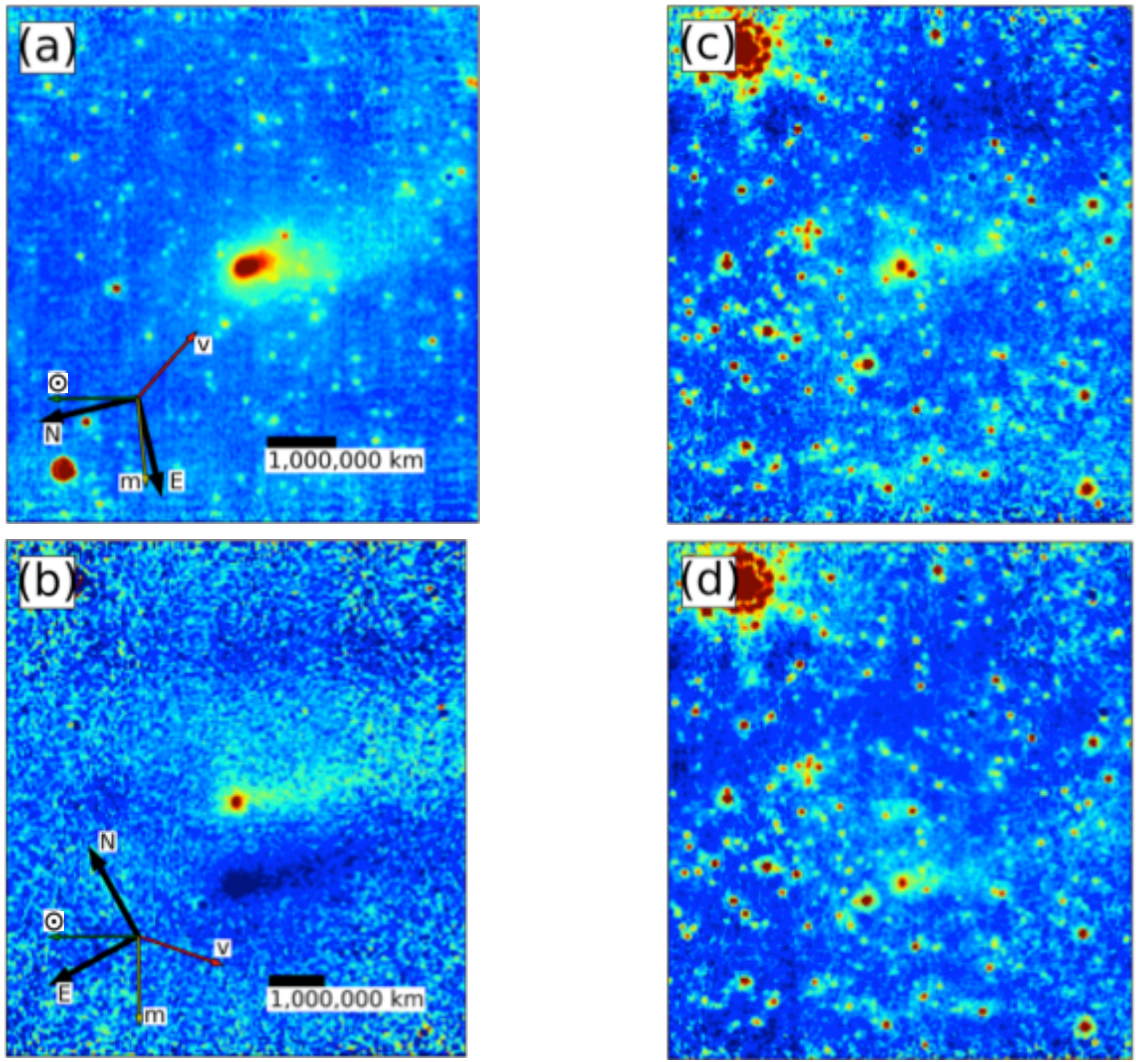

**Figure 1:** Mosaic images created from the 2005 and 2008 Spitzer data. N is north, E is east, ⊙ is the sunward direction, v is the heliocentric orbit velocity, and m is the apparent motion. The black bar represents $1.0*10^6$ km (projected) at the comet's physical location. The field of view shown here is 8.04 by 7.15 arcminutes. (a) 2005 data, one pixel represents ~38,000 km; (b)-(d) 2008 data, one pixel represents ~47,000 km. (b) shadow image of the two nights; (c) data from 31 August 2008; (d) data from 01 September 2008.

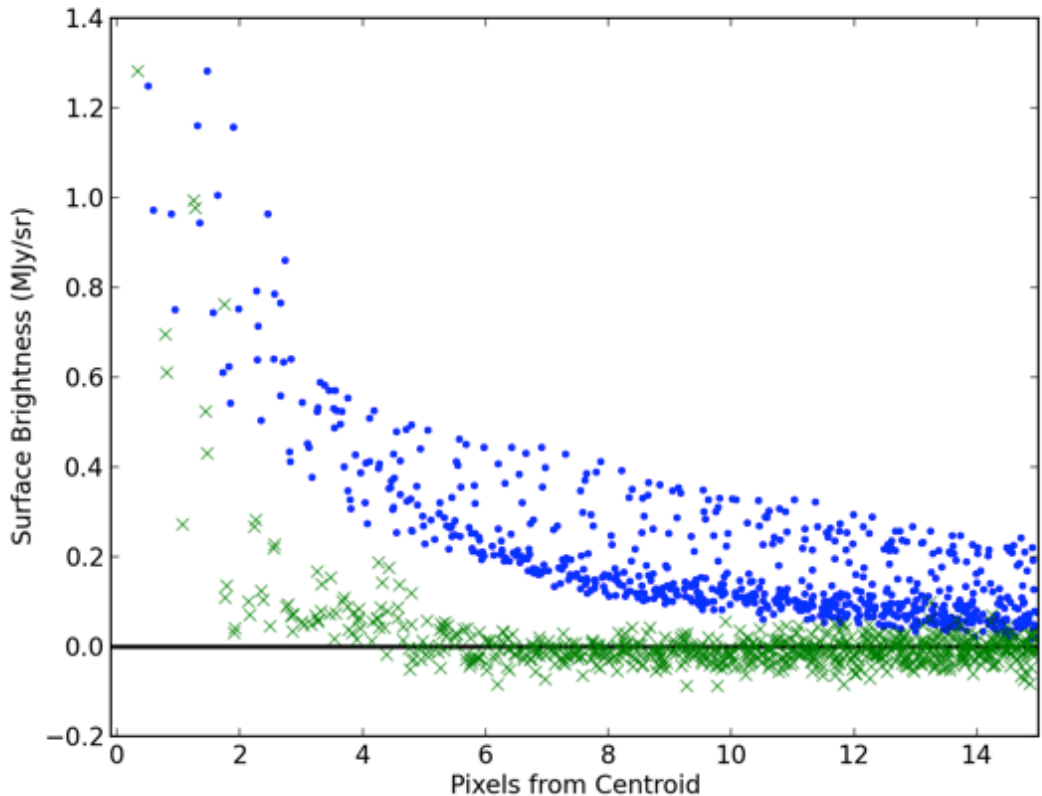

**Figure 2:** The radial profile of Hale-Bopp (blue o) compared to a field star (green x) in the 2005 data. It is clear that Hale-Bopp has a wider radial profile than the field star, suggesting that it has extended emission (i.e. a coma). Additionally, the peak brightness of the comet is offset from the centroid location, suggesting that the coma is asymmetric. The vertical scale of the star has been stretched to match that of Hale-Bopp, and a background level has been removed.

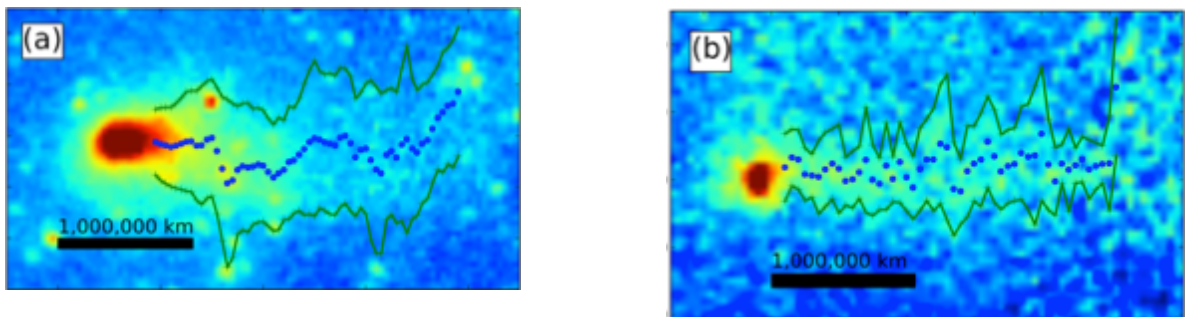

**Figure 3:** Gaussian fits to the tails. Blue dots are the Gaussian center, and the green lines represent one sigma width from the Gaussian center. The black bar represents a projected distance of 1,000,000 km at the comet's location. (a) 19 May 2005 data; (b) 31 August 2008 data.

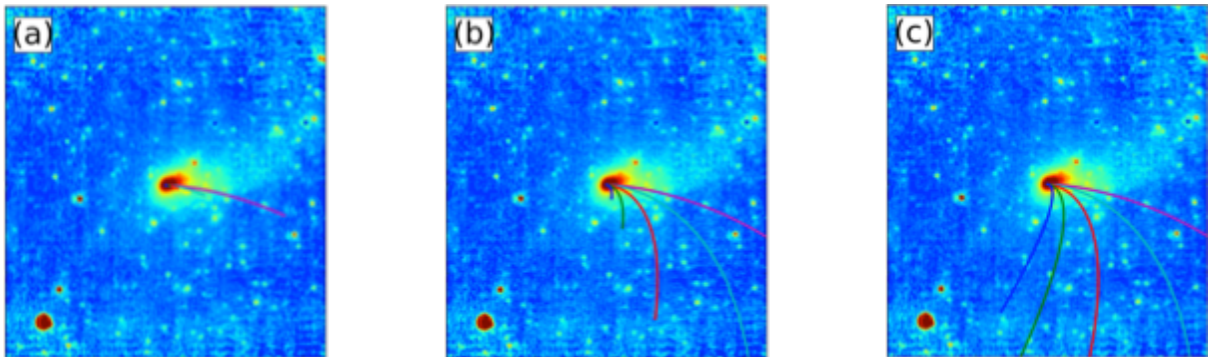

**Figure 4:** Syndynes generated using the Finson-Probstein models for the 2005 data. In each image, pink represents β=10, cyan is β=3, red is β=1, green is β=0.3 and blue is β=0.1. Maximum grain ages are (a) 1 year; (b) 3 years; (c) 5 years.

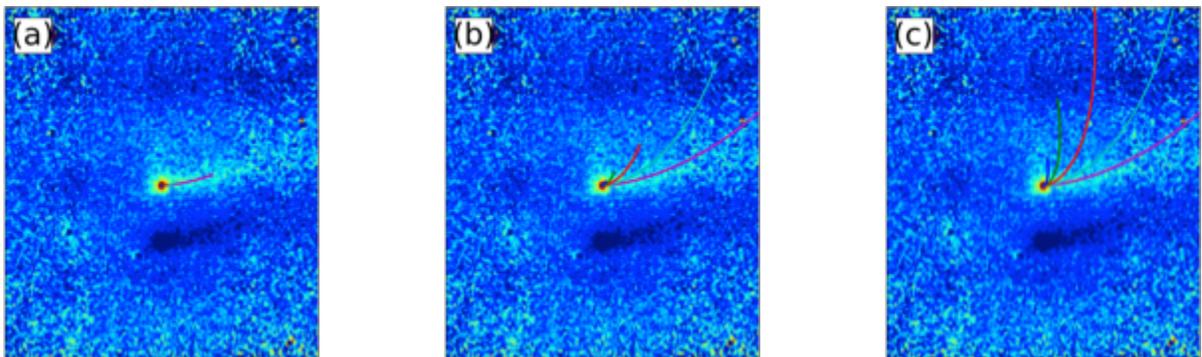

**Figure 5:** Syndynes generated using the Finson-Probstein models for the 2008 data. In each image, pink represents β=10, cyan is β=3, red is β=1, green is β=0.3 and blue is β=0.1. Maximum grain ages are (a) 1 year; (b) 3 years; (c) 5 years.

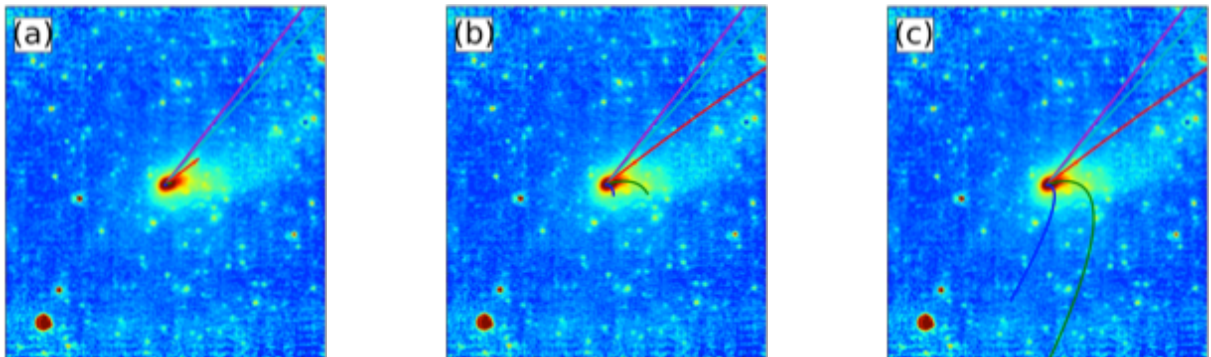

**Figure 6:** Syndynes generated using the Lorentz force model for the 2005 data. In each image, pink represents β=10 with q/m = 16.0, cyan is β=3 with q/m =14.4, red is β=1 with q/m =1.60, green is β=0.3 with q/m = 0.144 and blue is β=0.1 with q/m = 0.0160. Maximum grain ages are (a) 1 year; (b) 3 years; (c) 5 years.

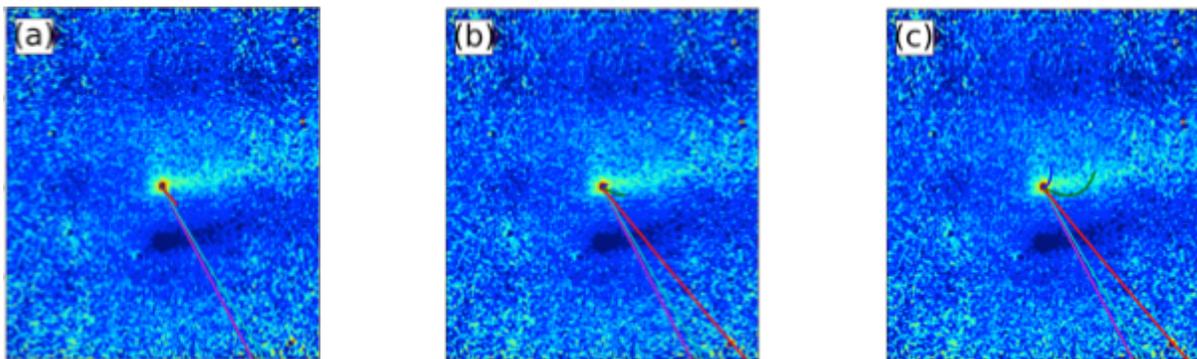

**Figure 7:** Syndynes generated using the Lorentz force model for the 2008 data. In each image, pink represents β=10 with *q/m* = 16.0, cyan is β=3 with q/m =14.4, red is β=1 with q/m =1.60, green is β=0.3 with q/m = 0.144 and blue is β=0.1 with q/m = 0.0160. Maximum grain ages are (a) 1 year; (b) 3 years; (c) 5 years.

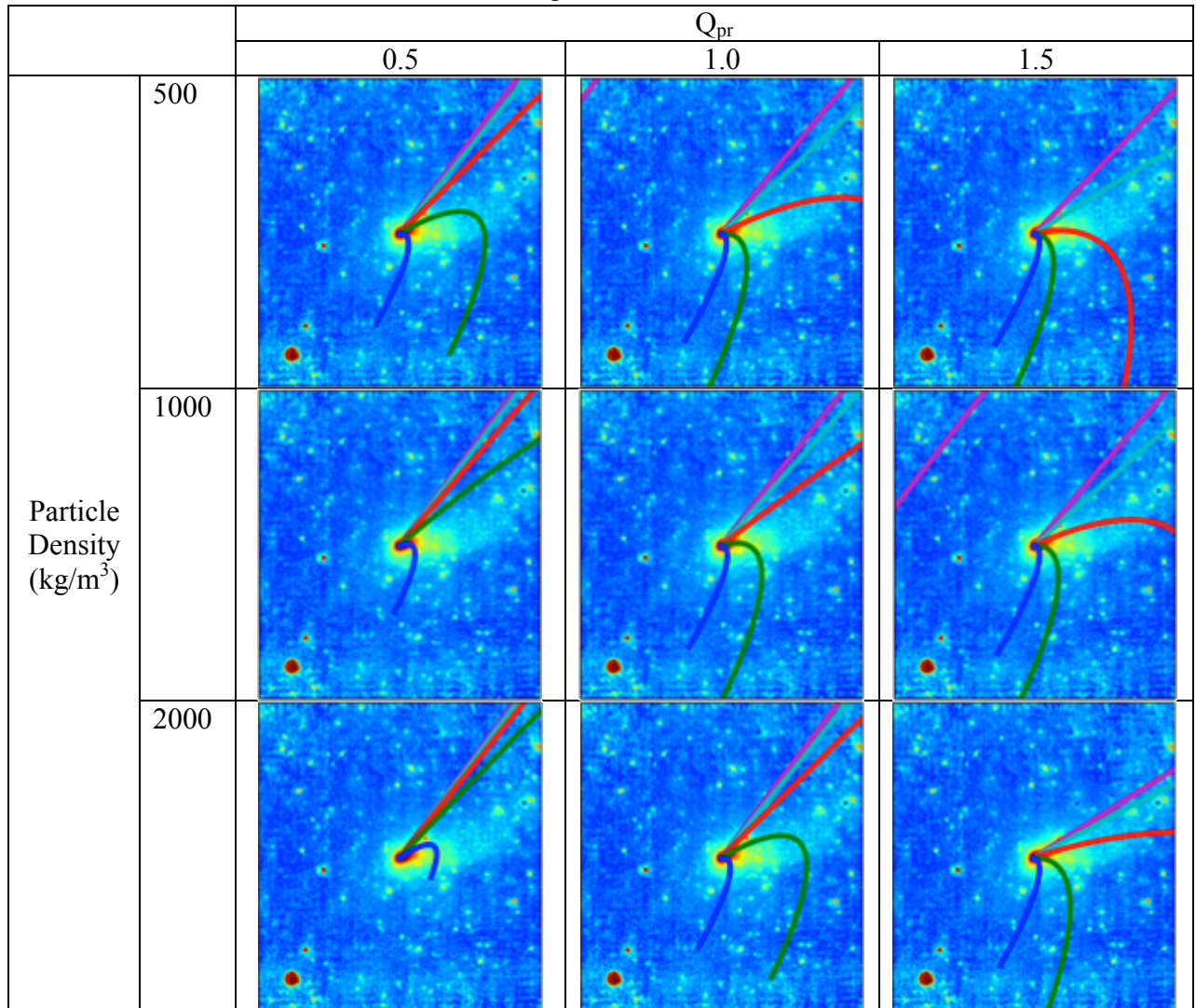

**Figure 8:** 5-year Lorentz force models for the 2005 data with a range of scattering efficiency ($Q_{pr}$) and particle density values. In each image, pink represents $\beta=10$, cyan is $\beta=3$, red is $\beta=1$, green is $\beta=0.3$, and blue is $\beta=0.1$.

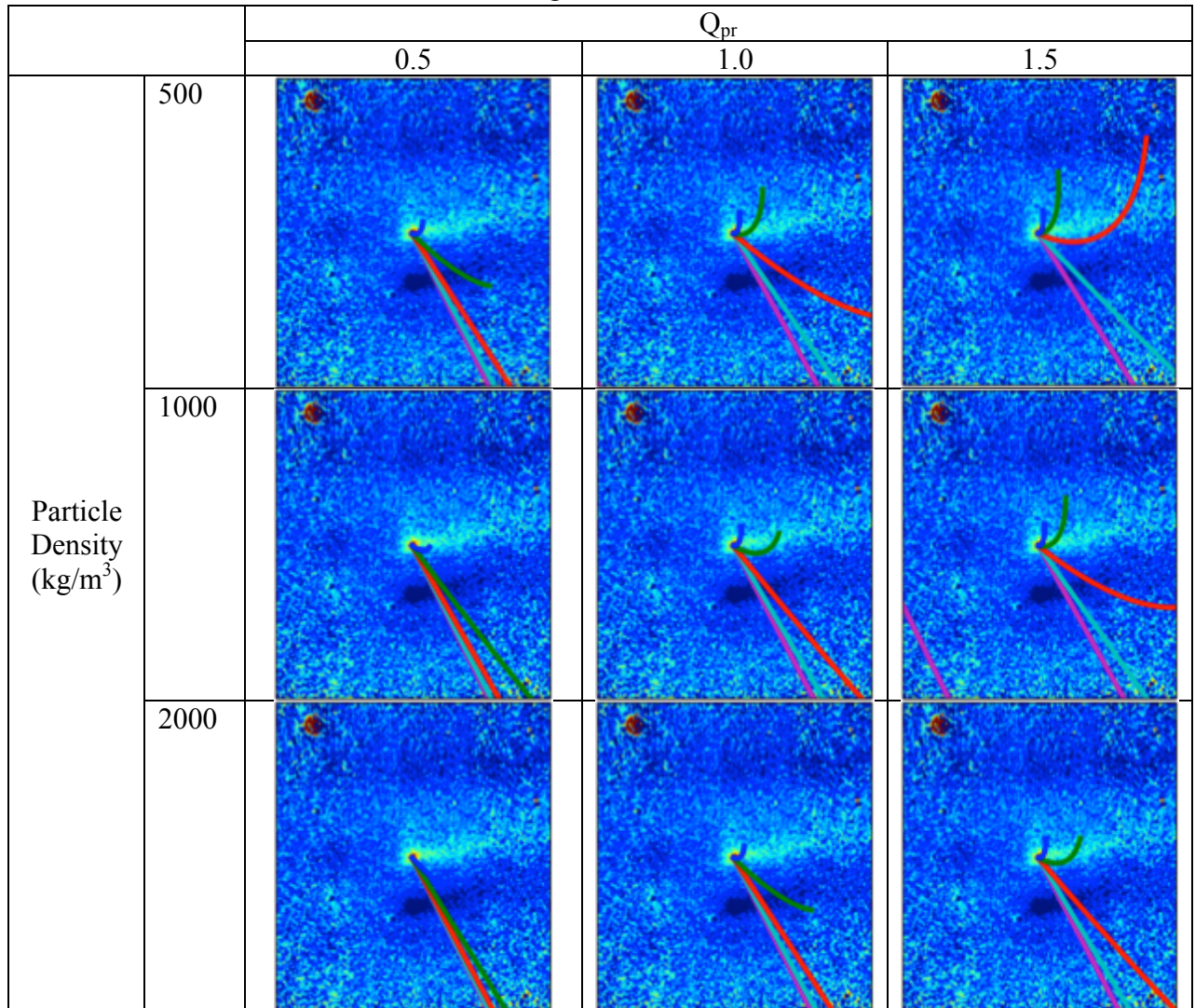

**Figure 9:** 5-year Lorentz force models for the 2008 data with a range of scattering efficiency ($Q_{pr}$) and particle density values. In each image, pink represents $\beta=10$, cyan is $\beta=3$, red is $\beta=1$, green is $\beta=0.3$, and blue is $\beta=0.1$.

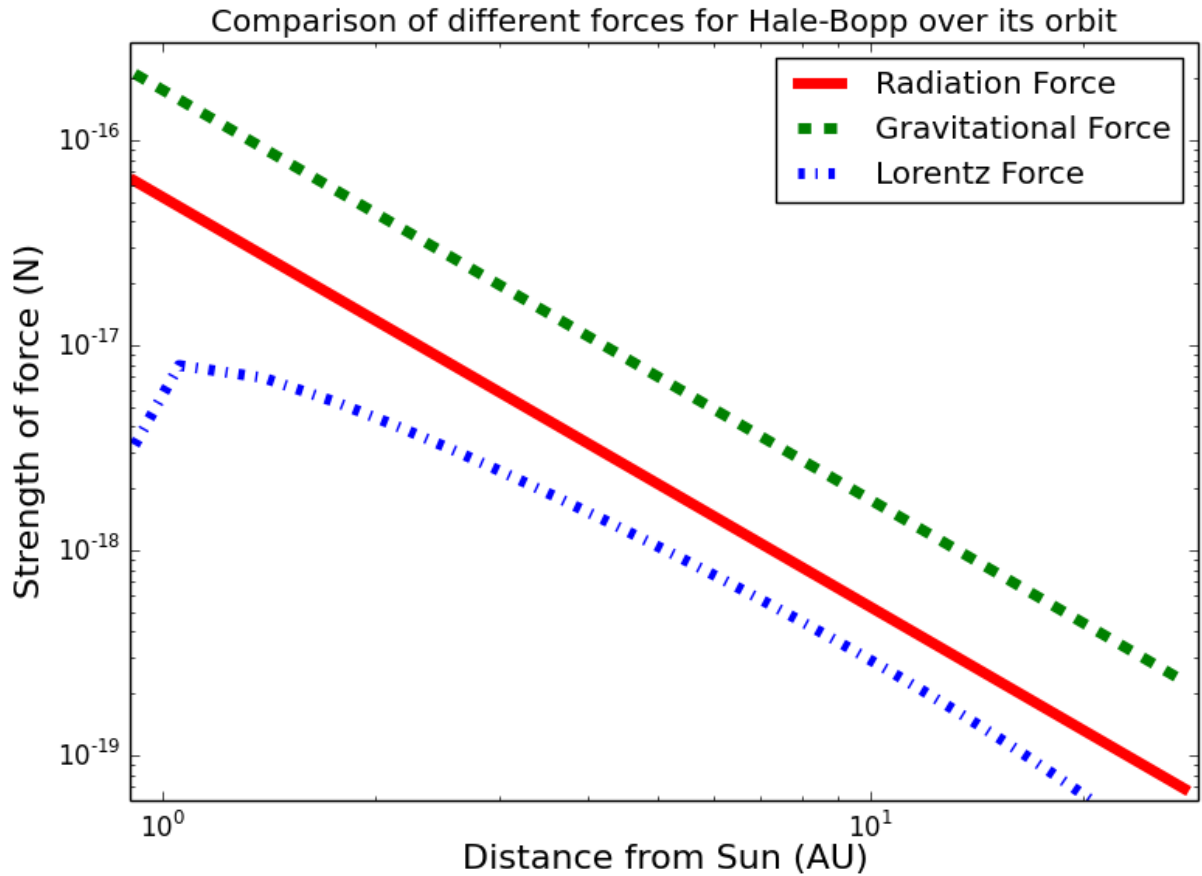

**Figure 10:** Comparison of the different forces on Hale-Bopp as function of its orbit. It is clear from this figure that the radiation force and gravitational force decrease as $1/r^2$, while the Lorentz force decreases as $(1/r)* \cos(\beta_{hg})$.

as revealed by Spitzer Space Telescope. *Icarus*, Volume 203, Issue 2, p. 571-588.